\documentclass{aastex62}
\received{ }
\revised{ }
\accepted{ }
\submitjournal{ApJ}

\shorttitle{Sample article}
\shortauthors{Luo et al.}
\begin{document}

\title{Hot Subdwarf Stars Identified in Gaia DR2 with Spectra of LAMOST DR6 and DR7. II.
Kinematics}

\correspondingauthor{Yangping Luo}
\email{ypluo@bao.ac.cn}

\author{Yangping Luo}
\affiliation{Department of Astronomy, China West Normal University, \\
Nanchong, 637002, PR China}
\author{P\'{e}ter~N\'{e}meth}
\affiliation{Astronomical Institute of the Czech Academy of Sciences, Fri\v{c}ova 298, CZ-251\,65 Ond\v{r}ejov, Czech Republic}
\affiliation{Astroserver.org, 8533 Malomsok, Hungary}

\author{Qida Li}
\affiliation{Department of Astronomy, China West Normal University, \\
Nanchong, 637002, PR China}

\begin{abstract}
 Combining the LAMOST radial velocities with Gaia parallaxes and proper motions, we presented 3D Galactic space motions and the orbits of 182 single-lined hot subdwarf stars. These stars have been identified by \citet{2020ApJ...889..117L} in Gaia DR2 with LAMOST DR6 and DR7 spectra. He-rich hot subdwarf stars with $\log(y)\ge0$ show the largest standard deviations of the Galactic velocity components and orbital parameters, while those with $-1\le\log(y)<0$ exhibit the second largest standard deviations. The two groups of He-deficient stars with $\log(y)<-1$ show similar standard deviations, which is systematically lower compared to He-rich stars. We also presented a kinematic population classification of the four hot subdwarf helium groups based on their positions in the $U-V$ velocity diagram, $J_{Z}-$eccentricity diagram and their Galactic orbits. The overall tendency of the fractional distributions of the four hot subdwarf helium groups in the halo, thin disk and thick disk is largely consistent with the findings reported by \cite{2019ApJ...881....7L} based on LAMOST DR5, which appears to support the predictions of binary population synthesis\citep{2003MNRAS.341..669H,2008A&A...484L..31H}. He-deficient stars with $-2.2\le\log(y)<-1$ likely origin from stable the Roche lobe overflow channel, He-deficient stars with $\log(y)<-2.2$  from the common envelope ejection channel, and He-rich stars with $\log(y)\ge0$ from the merger channel of double He white dwarf stars. The fraction of He-rich hot subdwarf stars with $-1\le\log(y)<0$ in the thin disk and the halo are far higher than in the thick disk, which implies that these stars have different formation channels in the thin disk and in the halo.
\end{abstract}


\section{Introduction} \label{sec:intro}
Hot subdwarf stars have been discovered back in 1954 \citep{1954AJ.....59..322G}.
They are situated at the blueward extension of the horizontal branch (HB) in the Hertzsprung-Russell (HR) diagram, which is also called the extreme horizontal branch (EHB) \citep{2009ARA&A..47..211H, 2016PASP..128h2001H}. Depending on their spectral appearance, hot subdwarf stars were traditionally classified as O type subdwarf (sdO) and B type subdwarf (sdB) stars \citep{2013A&A...551A..31D}.
They are core helium-burning stars with masses around $0.5M_{\odot}$. In spite of showing similar spectral appearance to O and B main sequence (MS) stars, they turned out to be much smaller than MS stars and, hence, much less luminous.
Being a typical product of stellar evolution, hot subdwarf stars play a vital role in understanding the properties of old stellar populations. They are responsible for the phenomenon called $UV$ upturn or $UV$ excess in the spectra of elliptical galaxies and the bulges of spiral galaxies \citep{2007MNRAS.380.1098H} and dominate the horizontal branch
morphology of globular clusters \citep{2008A&A...484L..31H, 2013A&A...549A.145L, 2015MNRAS.449.2741L}.
Hot subdwarf stars are also relevant for cosmology, as some of them are candidate progenitors of type Ia Supernovae \citep{2009A&A...493.1081J, 2009MNRAS.395..847W, 2010A&A...515A..88W, 2013A&A...554A..54G, 2015Sci...347.1126G, 2018RAA....18...49W}.

Hot subdwarf stars themselves are peculiar in more than one respect. Several types of pulsating stars have been discovered among hot subdwarfs and these objects turned out to be perfect laboratories for asteroseismic studies \citep{2012A&A...539A..12F, 2014A&A...569A..15O, 2018ApJ...853...98Z,2019MNRAS.482..758S}. They display very peculiar element abundance patterns, which mark active diffusion processes in their atmospheres, that is also responsible for their low helium abundances \citep{2003A&A...400..939E, 2016PASP..128h2001H, 2018MNRAS.475.4728B}.
A few intermediate helium hot subdwarf stars exhibit high abundances of lead, zirconium, strontium and yttrium, up to $10\,000$ times the solar values \citep{2011MNRAS.412..363N,2013MNRAS.434.1920N,2017MNRAS.465.3101J, 2019MNRAS.489.1481J,2019A&A...630A.130D, 2020MNRAS.491..874N}.
A substantial number of hot subdwarf stars have invisible compact companions, i.e. a neutron star or black hole.
Such systems are potential gravitational wave sources, that might be resolved by future facilities, such as the Laser Interferometer Space Antenna (LISA) \citep{2018A&A...618A..14W,2019arXiv191207705W}.

However, the formation of hot subdwarf stars is not well understood yet. Their formation requires the progenitors to lose almost their entire hydrogen envelope after passing the red giant branch (RGB). The remaining hydrogen envelope has not enough mass to sustain a hydrogen-burning shell. The reason for the very high mass loss prior to or at the beginning of the helium core flash is still unclear. Different scenarios have been put forward to explain this huge mass loss. The high fraction of binaries amongst hot subdwarfs suggests that binary evolution involving common envelope (CE) ejection, stable Roche lobe overflow (RLOF) or the merger of double helium white dwarfs (HeWD) are the main formation channels \citep{1984ApJ...277..355W,2002MNRAS.336..449H,2003MNRAS.341..669H}. Population synthesis studies indicated that the first two channels are responsible mainly for sdB stars and the merger channel for He-rich sdO stars \citep{2002MNRAS.336..449H, 2008A&A...484L..31H, 2012MNRAS.419..452Z}.
In between the sdB and sdO classes both the late hot-flasher scenario \citep{1996ApJ...466..359D,2004A&A...415..313M,2008A&A...491..253M} and the merger of helium white dwarfs with low mass main sequence stars \citep{2017ApJ...835..242Z} have been suggested to explain the origin of intermediate helium-rich hot subdwarf stars.
Although both of these models can explain the observed properties of hot subdwarfs, none of them appears entirely satisfactory.

With the advent of the Gaia survey \citep{2018A&A...616A...1G,2018A&A...616A..10G,2018A&A...616A..11G} and new spectroscopic surveys like LAMOST \citep{2012RAA....12.1197C} (the Large Sky Area Multi-Object Fiber Spectroscopic Telescope, also named the "Guo Shou Jing" Telescope), new and much larger observational samples shed light onto the details of hot subdwarf formation. A total of 166 hot subdwarf stars were identified by \cite{2016ApJ...818..202L} from LAMOST DR1 spectra.
\cite{2018ApJ...868...70L} spectroscopically confirmed 294 new hot subdwarf stars in Gaia DR2 with LAMOST DR5 spectra.
Recently, \cite{2019A&A...621A..38G} published a catalogue of $39\,800$ hot subdwarf candidates selected from Gaia DR2.
We have already presented the spectral analyses of 892 non-composite spectra hot subdwarf stars and the kinematics of 747 stars of that catalog by combining LAMOST DR5 and Gaia DR2 data \citep{2019ApJ...881....7L}.
Most recently, \citet{2020ApJ...889..117L} published the spectroscopic properties of 182 single-lined spectra hot subdwarf stars selected from Gaia DR2 with spectra of LAMOST DR6 and DR7, without discussing their kinematics.

Because kinematics can put strong constraints on our understanding of hot subdwarf formation we supplement our previous study \citep{2019ApJ...881....7L} in this paper. We present the kinematics of the 182 single-lined spectra from \citet{2020ApJ...889..117L} by combining the radial velocities (RV) extracted from LAMOST spectra with the parallaxes and proper motions from Gaia DR2. In Section 2 we introduce the targets and available data sets and describe the calculations of Galactic space velocities. Orbital parameters are discussed in Section 3. In Section 4, we discuss the Galactic space distribution, space velocity distribution, orbits, population classification and selection biases for the hot subdwarf groups of different helium abundances.
Finally, we draw conclusions in Section 5.

\section{targets and Data} \label{sec:targ}
\subsection{Targets}
We analysed a sample of 182 single-lined hot subdwarf stars observed in Gaia DR2 and LAMOST DR6 and DR7 \citep{2020ApJ...889..117L}.
The sample included 89 sdB, 37 sdOB, 26 sdO, 24 He-sdOB, 3 He-sdO and 3 He-sdB stars. The surface temperature $T_{\rm eff}$, gravity $\log\,g$, helium abundance $y=n{\rm He}/n{\rm H}$ were also collected from Table 1 by \citet{2020ApJ...889..117L} and are shown in Table 1. As described in \citet{2019ApJ...881....7L}, these 182 stars can be divided into four groups based on their helium abundances. Generally, the stars were classified as He-rich and He-deficient with respect to the solar helium abundance $\log{y}=-1$. Furthermore, He-rich and He-deficient stars can also be independently divided into two groups via $\log(y)=0$ and $\log(y)=-2.2$.
The classification scheme could inherently associate these four helium groups with different formation channels in the $T_{\rm eff}-\log(y)$ diagram \citep{2012MNRAS.427.2180N, 2019ApJ...881....7L}.
As described by \citet{2012MNRAS.427.2180N}, composite spectrum binaries with F and G type companions are relatively easy to identify because they have characteristic features, very different from subdwarfs and a comparable optical brightness. Identifying composite spectra with late G and K type companions is a challenge because of their significantly lower contributions and weaker lines. For these reasons, the identification of composite spectra with late type companions also heavily depends on the quality of the spectra. We excluded double-lined composite spectrum systems with noticeable Ca\,{\sc ii} H\&K  ($\lambda3933\AA$ and $\lambda3968\AA$), Mg\,{\sc i} ($\lambda5183\AA$), or Ca\,{\sc ii} ($\lambda8650\AA$) absorption lines. Unfortunately, the near infrared region is seriously polluted by sky emission lines in LAMOST spectra and we could not use the Ca\,{\sc ii} triplet lines.

Binary systems affect the calculations of Galactic velocities and orbits. Although our study focuses on studying only single-lined hot subdwarf stars, we cannot exclude the possibility of having unknown and unresolved binary systems based on a single epoch radial velocity measurement. We consider all stars to be members of the thin disk, thick disk or halo populations until they are further constrained.

\subsection{Data}
We utilised the spectra of LAMOST DR6 and DR7 to measure the radial velocities of the 182 hot subdwarf stars. The LAMOST spectra are similar to the SDSS data having the resolution $R\thicksim1800$ and wavelength coverage of $3800-9100$ \AA. Further details of the data have been described in \cite{2012RAA....12.1243L,2014IAUS..298..428L}. The published radial velocities in the LAMOST catalog are not reliable for hot subdwarf stars, because hot subdwarfs are not included in LAMOST stellar templates for RVs. Therefore, we re-measured the radial velocities of these 182 stars and present them in Table \ref{tab:tab1}.

Gaia DR2 provided high-precision positions ($\alpha$ and $\delta$), proper motions ($\mu_{\alpha}\cos\delta$ and $\mu_{\delta}$) and parallaxes ($\bar{\omega}$) \citep{2018A&A...616A...1G,2018A&A...616A..10G, 2018A&A...616A..11G} for all 182 stars. Distances ($D$) were calculated by using $D=1/\bar{\omega}$. These parameters are shown in Table \ref{tab:tab1}. However, for 20 stars reliable distances cannot be obtained by simply inverting the parallax. Therefore, their distances were replaced with the estimated values from the Gaia-DR2 distance catalogue \citep{2018AJ....156...58B}.

\section{Galactic space velocities and orbital parameters}\label{sec:cal}
Based on the distances, radial velocities and proper motions shown in Table \ref{tab:tab1}, we calculated space velocity components in the Cartesian coordinates with the \textit{Astropy} Python package. We adopted a right-handed Galactocentric Cartesian coordinate system, where the velocity components $U$, $V$, and $W$ are positive in the direction towards the Galactic center, Galactic rotation and north Galactic pole, respectively. We set the distance of the Sun from the Galactic centre to be 8.4 kpc and the velocity of the local standard of rest (LSR) to be $242\,{\rm km\,s^{-1}}$ \citep{2013A&A...549A.137I}. For the solar velocity components with respect to the LSR, we assumed ($U_{\odot}$, $V_{\odot}$, $W_{\odot}$)$=$(11.1, 12.24, 7.25)$\,{\rm km\,s^{-1}}$ \citep{2010MNRAS.403.1829S}. Making use of \textit{Astropy}, we also computed the space position components in a right-handed Galactocentric Cartesian reference frame denoted by ($X$, $Y$, $Z$).

We applied the \textit{Galpy} Python package \citep{2015ApJS..216...29B} to calculate the Galactic orbital parameters of our program stars. For the calculation of orbits, we adopted the Milky Way potential "MWpotential2014" that comprises a power-law bulge with an exponential cut-off, an exponential disk and a power-law halo component \citep{2015ApJS..216...29B}. We used the same solar Galactocentric distance and LSR velocity as in \textit{Astropy}. The Galactic orbital parameters of hot subdwarf stars, such as the apocentre ($R_{\rm ap}$), pericentre ($R_{\rm peri}$), eccentricity ($e$), maximum vertical amplitude ($z_{\rm max}$), normalised z-extent ($z_{n}$) and z-component of the angular momentum ($J_{z}$), are extracted from integrating their orbital paths for a time of 5 Gyrs and are listed in Table$\,$\ref{tab:tab1}. $R_{\rm ap}$ and $R_{\rm peri}$ represent the maximum and minimum distances of an orbit from the Galactic center, respectively.
We defined the eccentricity by
\begin{equation}
e=\frac{R_{\rm ap}-R_{\rm peri}}{R_{\rm ap}+R_{\rm peri}},
\end{equation}
and the normalised z-extent by
\begin{equation}
z_{n}=\frac{z_{\rm max}}{R(z_{\rm max})},
\end{equation}
where $R$ is the Galactocentric distance.

The errors of the space positions and velocity components, as well as of the orbital parameters were obtained with a Monte Carlo simulation. For each star $1\,000$ sets of input values with a Gaussian distribution were simultaneously generated and the output parameters were computed together with their errors. Further details on the calculations can be found in \citet{2019ApJ...881....7L, 2020NewA...7801363L}.

\section{Results} \label{sec:res}
\subsection{Space distribution} \label{subsec:spac}
\begin{figure*}
\epsscale{1.2}
\plotone{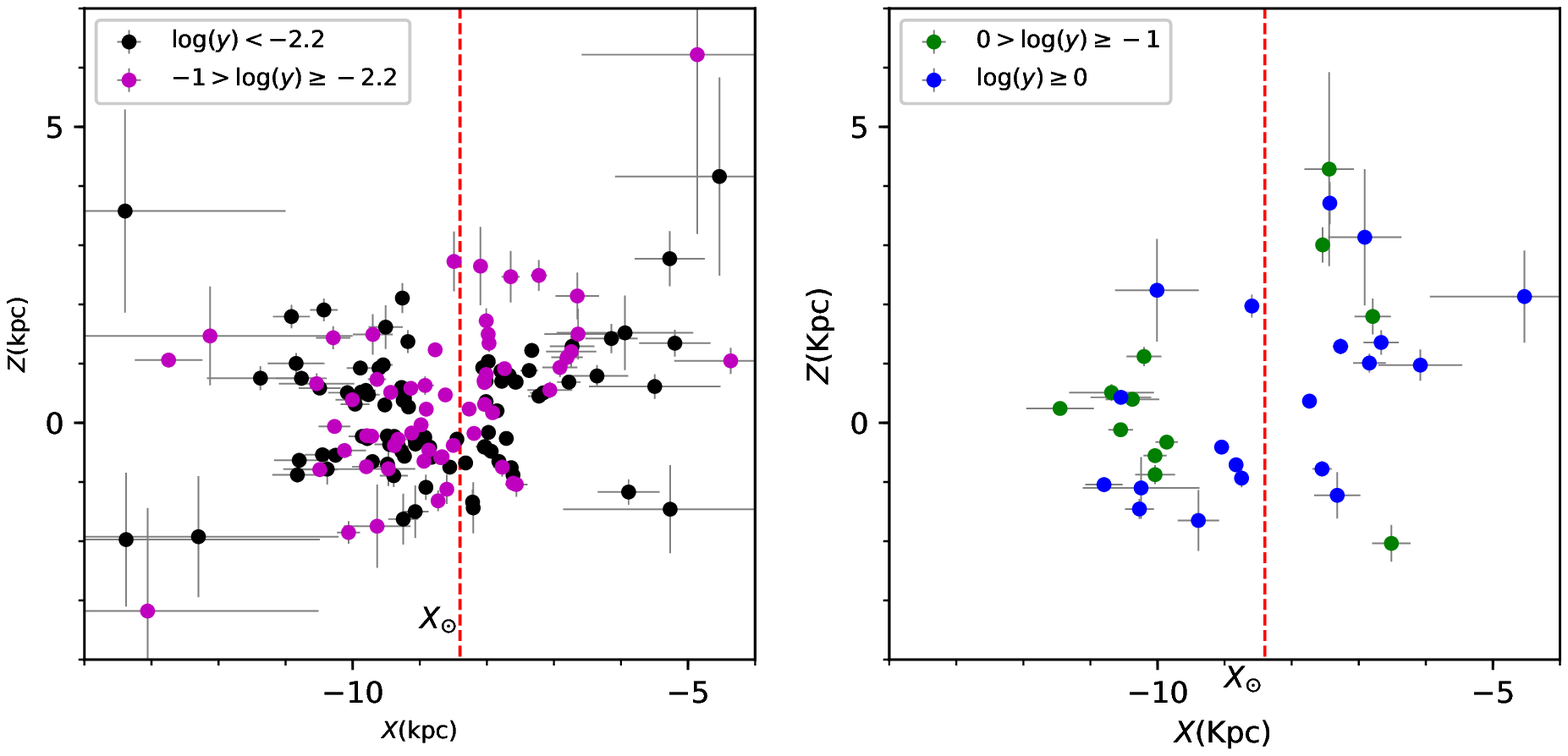}
\caption{The space positions of hot subdwarf stars in Cartesian Galactic $X-Z$ coordinates. He-deficient stars are shown in the left panel and He-rich stars are displayed in the right panel. The dashed line marks the solar position. \label{fig:fig1}}
\end{figure*}

Fig$\,$\ref{fig:fig1} displays the space positions of the four hot subdwarf helium groups in the $X-Z$ diagrams.
The left panel of Fig$\,$\ref{fig:fig1} reveals that the space distributions of the two He-deficient groups do not show any obvious differences. Most stars tend to cluster around the disk and only a few stars are found in the halo. The star density quickly decreases from the disk to the halo and a sharp cut-off appears at $|Z|\sim1.5\,{\rm kpc}$, which is considered as the vertical scale height of the thick disk \citep{2017MNRAS.467.2430M}.

In contrast, the right panel of Fig$\,$\ref{fig:fig1} exhibits that the space distributions of the two groups of He-rich stars have a noticeable difference at $|Z|>1.5\,{\rm kpc}$ where the star density of He-rich stars with $\log(y)\ge0$ is significantly higher than that of stars with $-1\le\log(y)<0$. The difference in space distribution also indicates that the two groups of He-rich stars are likely origin from different formation channels.

Comparisons of the left and right panel in the Fig$\,$\ref{fig:fig1} demonstrate that the space density of the two group of He-rich stars has a larger dispersion than the groups of He-deficient stars, which suggests that He-rich and He-deficient hot subdwarf stars have different kinematic origins.

\subsection{Galactic velocity distribution} \label{subsec:cal}
\begin{figure*}
\epsscale{0.9}
\plotone{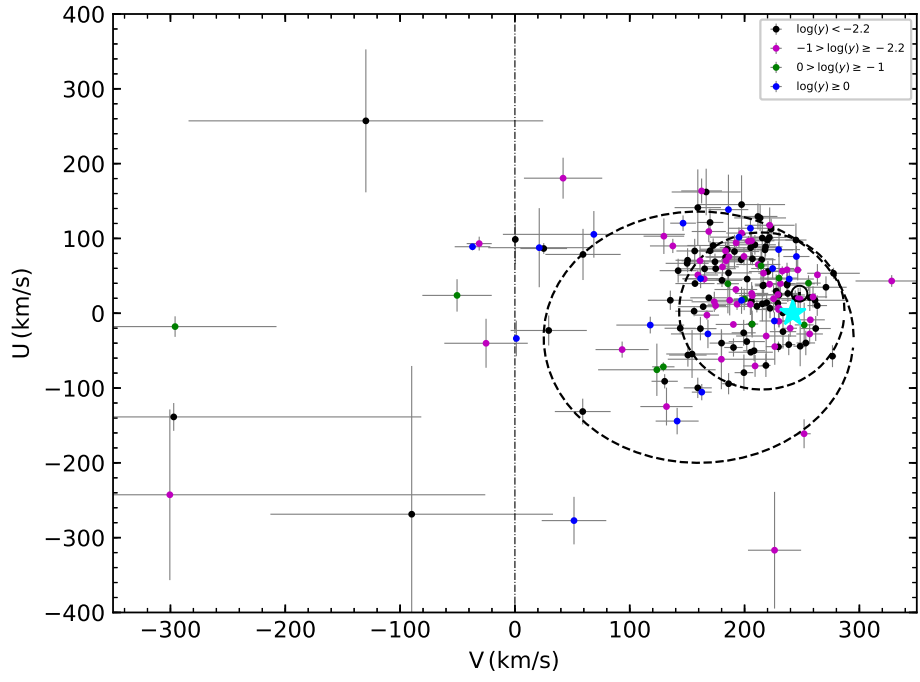}
\caption{$U-V$ velocity diagram for the four hot subdwarf helium groups. Two dashed ellipses denotes the $3\sigma$ limits for the thin disk and thick disk populations, respectively \citep{2006A&A...447..173P}. The cyan star symbol represents the Local Standard of Rest (LSR).
\label{fig:fig2}}
\end{figure*}

Fig$\,$\ref{fig:fig2} exhibits the distribution of the four hot subdwarf helium groups in the $U-V$ velocity diagram. The $U-V$ velocity diagram demonstrates that He-deficient stars can be found mostly around the LSR, while He-rich stars are more widely scattered in the whole region. In order to identify the Galactic population memberships of the stars, we also plot the two dotted ellipses as shown in Fig$\,$1 by \cite{2017MNRAS.467...68M}. They mark the $3\sigma-$limits of thin and thick disk WDs \citep{2006A&A...447..173P}, respectively.

\begin{figure*}
\epsscale{0.85}
\plotone{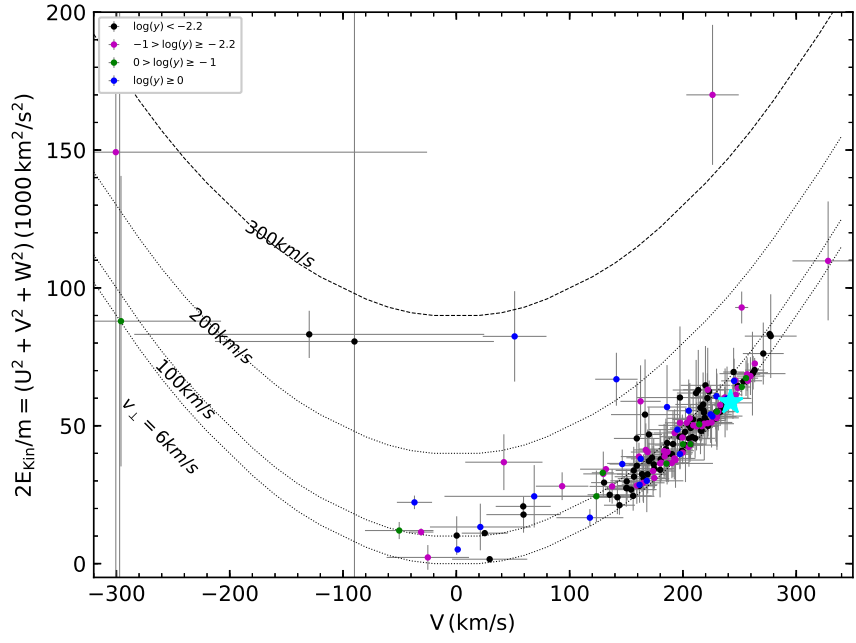}
\caption{Galactic rotational velocity $V$ against the total kinetic energy $2E_{\rm kin}/m=U^{2}+V^{2}+W^{2}$ for the four hot subdwarf helium groups. The parabolic curves denote the isovelocity perpendicular to the direction of Galactic rotation, where $V_{\perp}=(U^{2}+V^{2})^{1/2}$. The cyan star symbol represents the Local Standard of Rest (LSR).
\label{fig:fig3}}
\end{figure*}

In order to look at the kinematics of the total velocity for the four hot subdwarf helium groups, Fig$\,$\ref{fig:fig3} displays the kinetic energy $2E_{\rm kin}/m=U^{2}+V^{2}+W^{2}$ versus rotational velocity ($V$) diagram. The higher the value of the kinetic energy $2E_{\rm kin}/m$, the more elliptic is the orbit of the star. We also plotted the isovelocity curves perpendicular to the Galactic rotation, where $V_{\perp}=(U^{2}+V^{2})^{1/2}$. The higher the value of the $V_{\perp}$, the hotter is the kinematic temperature. As described by \cite{2019ApJ...881....7L}, most of stars are clustered around the LSR in a "banana" shaped region alongside the $V_{\perp}=0{\rm \,km\,s^{-1}}$ isovelocity curve, which means that they are kinematically cool and likely have more circular orbits. A few stars are located further away from the $V_{\perp}=0 {\rm \,km\,s^{-1}}$ isovelocity curve where He-rich stars with $\log(y)\ge0$ have a higher fraction. These are kinematically hot stars with likely more eccentric orbits.
The sample also exhibits a sharp cut near $110{\rm \,km\,s^{-1}}$. The few stars to the left of this velocity limit show a larger scatter and belong to the halo population \citep{2004A&A...414..181A}. In this region, the proportion of He-rich stars with $\log(y)\ge0$ is more than $25\%$.

Table\,\ref{tab:tab2} lists the mean values and standard deviations of the Galactic velocity components for the four hot subdwarf helium groups. We find that He-rich stars with $\log(y)\ge0$ show the largest standard deviations of the Galactic velocity components in all four hot subdwarf helium groups and He-rich stars with $-1\le\log(y)<0$ display the second largest standard deviation. The two groups of He-deficient stars exhibit similar values of standard deviations.

These results are in good agreement with the findings of \cite{2019ApJ...881....7L}. The diverse range of kinematic velocities support that He-rich hot subdwarf stars with $\log(y)\ge0$ are likely originate from different formation channels.

\subsection{Galactic orbits} \label{sec:orb}
\begin{figure*}
\epsscale{0.9}
\plotone{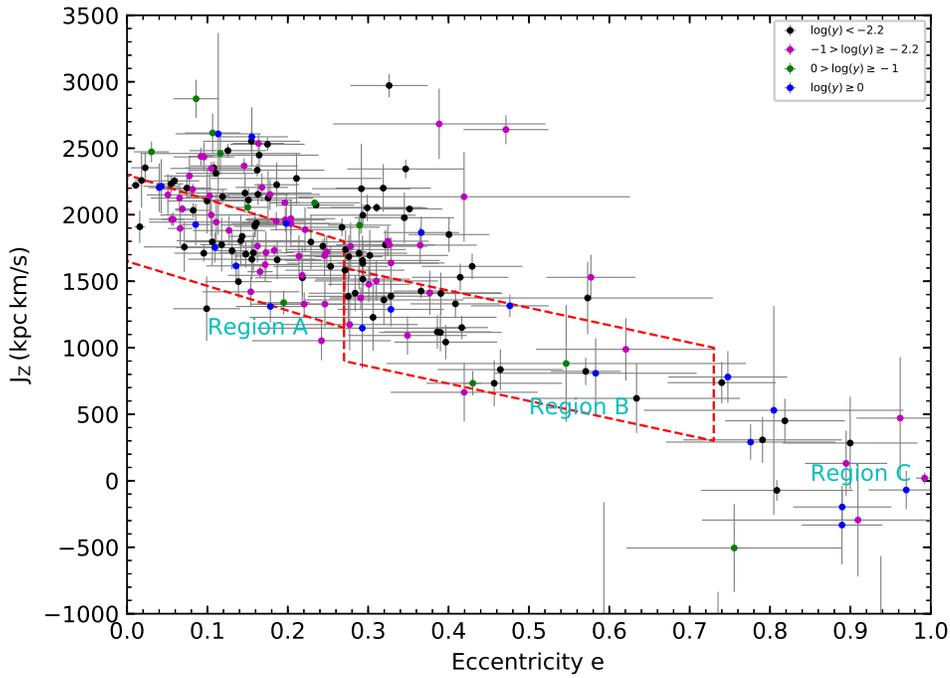}
\caption{Z-component of the angular momentum versus eccentricity ($e$) for the four hot subdwarf helium groups. The two parallelograms denote Region A (thin disk) and Region B (thick disk) \citep{2006A&A...447..173P}.
\label{fig:fig4}}
\end{figure*}

Two important orbital parameters are the z-component of the angular momentum $J_{z}$ and the eccentricity $e$ of the orbit. They are used to distinguish different populations. Fig$\,$\ref{fig:fig4} shows the distribution of the four hot subdwarf helium groups in the $J_{z}-e$ diagram.
We also show the two regions defined by \cite{2003A&A...400..877P}: Region A confines thin disk stars clustering in an area of low eccentricity, and $J_{z}$ around $1\,800\,{\rm kpc\,km\,s^{-1}}$. Region B encompasses thick disk stars having higher eccentricities and lower angular momenta. Outside of these two regions, defined as region C, halo star candidates are found. The majority of stars show a continuous distribution from Region A to Region B without an obvious dichotomy. Only a few stars lie in Region C and they they are separated by a noticeable gap from Region B and Region C. In Region C, He-rich stars with $\log(y)\ge0$ have a very high fraction.

In Table \ref{tab:tab2} we give the mean values and standard deviations of the orbital parameters: eccentricity, normalised z-extent, maximum vertical amplitude, apocentre and pericentre. The standard deviation of the orbital parameters are similar to that of the Galactic velocity components.
He-rich stars with $\log(y)\ge0$ show the largest standard deviation of the orbital parameters and He-rich stars with $-1\le\log(y)<0$ display the second largest standard deviation. The two groups of He-deficient stars show similar values of standard deviations.
These results are in good agreement with earlier findings \citep{2017MNRAS.467...68M, 2019ApJ...881....7L}.

\subsection{Galactic Population classifications}\label{subsec:pop}
We primarily adopted the $U-V$ diagram, $J_{z}-e$ diagram and the maximum vertical amplitude $z_{\rm max}$ to distinguish the Galactic populations of hot subdwarfs. To ensure correct population assignments, all orbits were visually inspected to supplement the automatic classifications.
The detailed classification scheme was described by \cite{2017MNRAS.467...68M} and \cite{2019ApJ...881....7L}. Thin disk stars are situated within the $3\sigma$ thin disk contour in the $U-V$ diagram and Region A in the $J_{z}-e$ diagram. Their orbits show a small extension in the Galactocentric distance $R$ and the Galactic plane $Z$ directions and have $z_{\rm max}<1.5\,{\rm kpc}$. Thick disk stars lie within the $3\sigma$ thick disk contour and in Region B. The extension of their orbits in the $R$ and the $Z$ directions are larger than that of thin disk stars, but do not reach the region of halo stars. Halo stars lie outside Region A and B as well as outside the $3\sigma$ thick disk contour. Their orbits show high extensions in $R$ and $Z$. There are also some halo stars with an extension in $R$ larger than $18 \rm{kpc}$, or the vertical distance from the Galactic plane $Z$ larger than $6 \rm{kpc}$.
Table$\,$\ref{tab:tab2} gives the number of stars in the four hot subdwarf helium groups classified as halo, thin or thick disk stars and Fig$\,$\ref{fig:fig5} displays their fractions in the halo, thin disk and thick disk.

The general trends in the distributions of the four hot subdwarf helium groups observed in LAMOST DR6 and DR7 can be matched with the results reported in LAMOST DR5 \citep{2019ApJ...881....7L}. A study on the structure of the Milky Way \citep{2017ApJS..232....2X} demonstrated that the different Galactic populations (thin disk, thick disk and halo) reflect different age stellar populations. The binary population synthesis calculations of \cite{2008A&A...484L..31H} gave the fractions of hot subdwarf stars from three different formation channels (stable RLOF, CE ejection and the merger of double HeWDs) at various stellar population ages. Although the exact values of the fractions are not consistent with the predictions of binary population synthesis \citep{2003MNRAS.341..669H,2008A&A...484L..31H}, we could make a comparison of the overall tendency of the fractional distributions.

The frequency of He-rich hot subdwarf stars with $\log(y)\ge0$ monotonically increases from $6\%$ in the thin disk to $23\%$ in the halo. This trend is in a good agreement with the predictions of the merger channel of double HeWDs. Although many observations could outline two groups of He-deficient stars in the $T_{\rm eff}-\log\,g$ and $T_{\rm eff}-\log(y)$ diagrams separated by a gap in He abundance at $\log(y)=-2.2$ \citep{2003A&A...400..939E, 2005A&A...430..223L, 2007A&A...462..269S, 2009PhDT.......273H, 2012MNRAS.427.2180N, 2011A&A...530A..28G, 2015A&A...577A..26G, 2016ApJ...818..202L,2019ApJ...881....7L, 2018ApJ...868...70L, 2020ApJ...889..117L}, their formation channels are not understood well.
\citet{2012MNRAS.427.2180N} found that hot subdwarf binary systems with F and G type companions, which are predominantly long-period binary candidates from the stable RLOF channel, appear in the two groups of He-deficient stars but show higher fractions among sdB stars with $-2.2 \le \log(y) < -1$.
However, reviews of larger samples \citep{2015MNRAS.450.3514K, 2015A&A...576A..44K} found that both short-period and long-period hot subdwarf binary systems occur in each sdB group.
We found that the fraction of He-deficient stars with $\log(y) < -2.2$  is in good agreement with the predictions of the CE ejection channel and the fraction of He-deficient sdB stars with $-2.2 \le \log(y) < -1$ agrees with the predictions of the stable RLOF channel if the excluded composite binary systems were all considered to have sdB stars with $-2.2 \le \log(y) < -1$ in LAMOST DR5. The vast majority of the identified composite spectra show signatures of F or early G type companions. To find the nature of hot subdwarfs in these systems  we will need spectral decomposition.
The distribution of single-lined He-deficient the hot subdwarf stars observed in LAMOST DR6 and DR7 \citep{2020ApJ...889..117L} is in good agreement with the distribution of single-lined He-deficient stars derived from LAMOST DR5 data \citep{2019ApJ...881....7L}.  These samples support the predicted fractional contributions of the formation channels \citep{2003MNRAS.341..669H,2008A&A...484L..31H}.

Finally, the formation of He-rich hot subdwarf stars with $-1\le\log(y)<0$ remains a puzzle.  The fraction of He-rich hot subdwarf stars with $-1\le\log(y)<0$ increases to $15\%$ in the halo after decreasing from $\sim8\%$ in the thin disk to $\sim3\%$ in the thick disk, which is consistent with that of LAMOST DR5. Their frequency implies that He-rich hot subdwarf stars with $-1\le\log(y)<0$ in the thin disk and the halo may have different formation channels.
Recent observations \citep{2017MNRAS.465.3101J,2019MNRAS.489.1481J,2019A&A...630A.130D,2020MNRAS.491..874N} found that He-rich hot subdwarf stars with $-1\le\log(y)<0$ show a strong enrichment of heavy elements. The reason of this enrichment is still unclear. Future kinematic studies may help shed light onto the poorly understood physical process behind the strong enrichment of heavy elements.
\subsection{\bf{Discussion of selection biases}}\label{subsec:dis}
\begin{figure*}
\epsscale{1.0}
\plotone{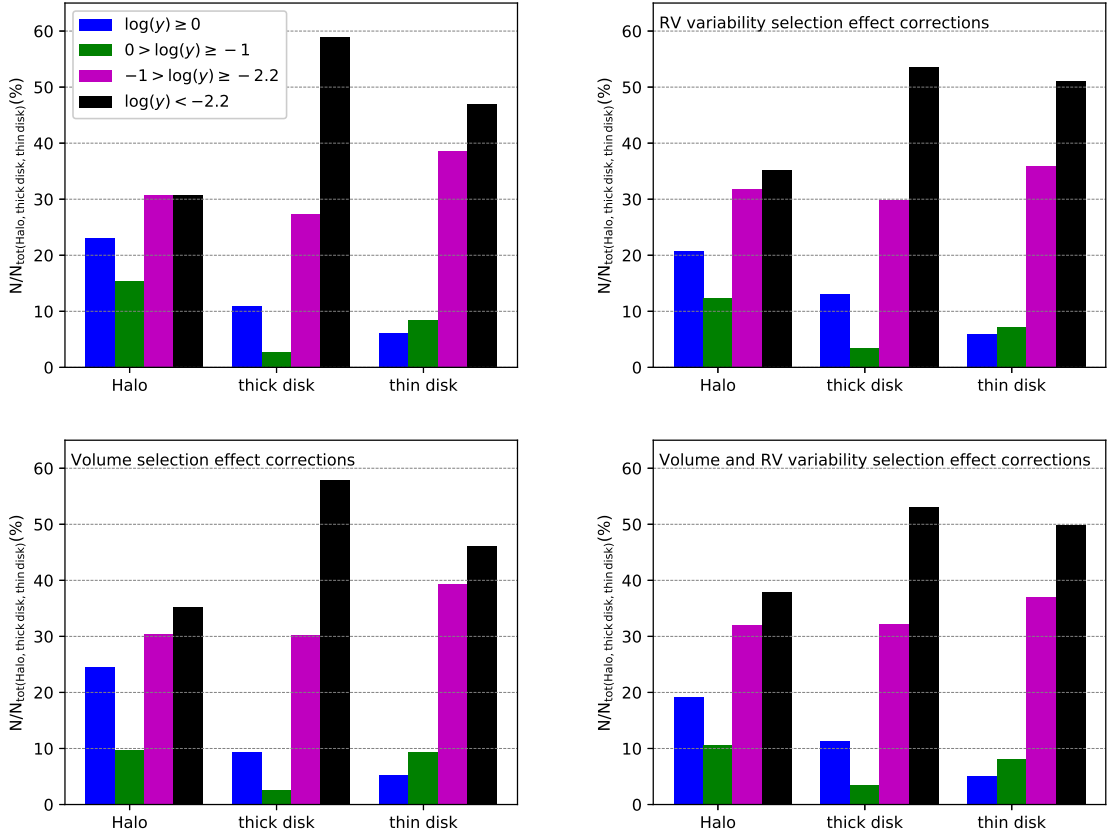}
\caption{The fractional distributions of the four hot subdwarf helium groups for the halo, thick disk and thin disk populations. Upper left: uncorrected. Upper right: RV variability selection effect corrections. Lower left: Volume selection effect corrections. Lower right: Volume and RV variability selection effect corrections.
\label{fig:fig5}}
\end{figure*}

 Radial velocity surveys (e.g., \citet{2001MNRAS.326.1391M, 2003MNRAS.338..752M, 2011MNRAS.415.1381C,2011A&A...526A..39G}) of sdB stars showed that about $50\%$ of sdB stars reside in close binary systems with either a cool MS star or a WD companion. \citet{2004Ap&SS.291..321N} reported a binary fraction of $39\%$ of sdB stars from the ESO Supernova type Ia Progenitor survey (SPY). Recently, \citet{2015MNRAS.450.3514K} reported a binary fraction of $37\%$ of hot subdwarf stars selected from the GALEX all-sky survey and showed RV amplitudes ranging from a few tens to hundreds of $km/s$. The kinematic analysis based on just one epoch in RV is therefore intrinsically uncertain. With the binary population statistics of \citet{2015MNRAS.450.3514K}, we performed Monte Carlo simulations for our sample. We applied the binary fraction of $37\%$ for single-lined subdwarf stars and the distribution of RV amplitudes to correct for systematics due to the unknown RV. For each binary system, we assumed a circular orbit in the form $RV(t)=\gamma+K\sin\phi$, where $K$ is the RV amplitude, $\gamma$ is the system velocity and $\phi$ is the orbital phase. The orbital phase $\phi$ was chosen from a uniform distribution from 0 to $2\pi$. 3000 system RVs were produced for each individual star. Combing the distances, proper motions and their errors, we calculated their Galactic space velocity components and orbits. We obtained the probabilities of the Galactic populations on each individual star and listed in Table \ref{tab:tab1}. The upper right panel of Figure$\,$\ref{fig:fig5} shows the RV variability selection effect corrected fractional distributions of the four hot subdwarf helium groups for the halo, thick disk and thin disk populations.  The impact of RV variability selection effect on the fractional distributions is less than $5\%$ of the number of stars in a group.

\begin{figure*}
\epsscale{0.8}
\plotone{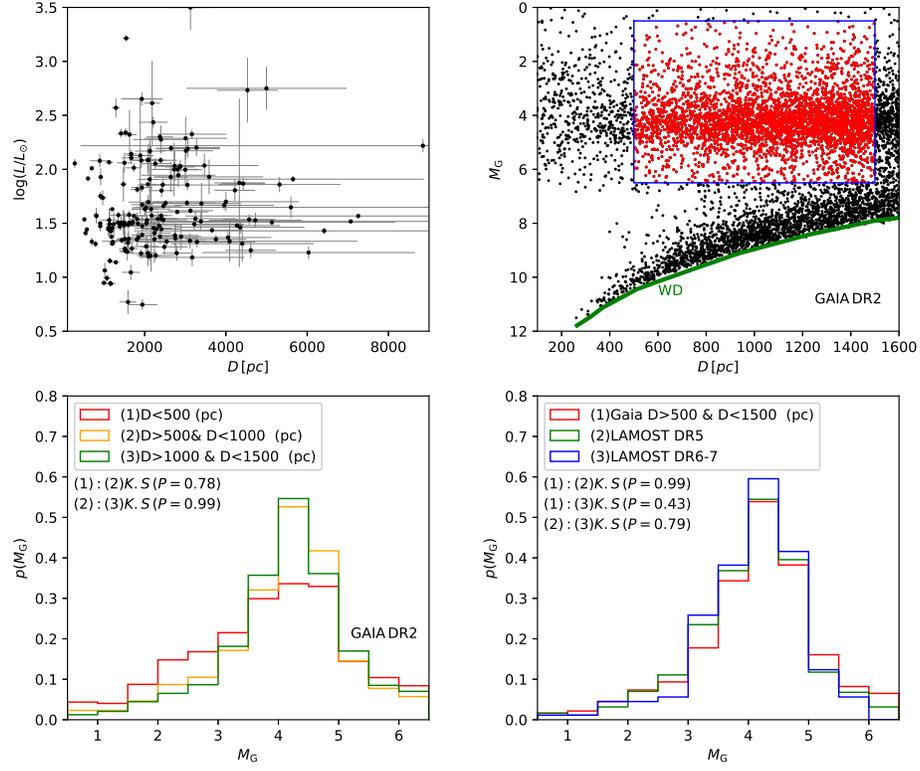}
\caption{Selection effect corrections. Upper left: luminosity (assuming a stellar mass of $0.47M_{\odot}$) versus Gaia distance for 182 hot subdwarf stars in LAMOST DR6 and DR7. Upper right: Gaia absolute G magnitude versus Gaia distance for hot subdwarf stars in Gaia DR2 \citep{2019A&A...621A..38G}. The green line denotes the cut-off value for WDs. Lower left: distribution functions of Gaia absolute G magnitude for three distance intervals for hot subdwarf stars in Gaia DR2. Lower right: comparison of the distribution functions of hot subdwarf stars in LAMOST DR5, DR6, DR7 and Gaia DR2.
\label{fig:fig6}}
\end{figure*}

Using the effective temperature ($T_{\rm eff}$) and surface gravity ($g$) we determined the total luminosity (in $L_{\odot}$) by assuming for all stars a sample-average mass of $0.47M_{\odot}$ \citep{2012A&A...539A..12F}. The upper left panel of Figure$\,$\ref{fig:fig6} displays the luminosity versus distance of the sample. There is no clear correlation between luminosity and distance for the sample.

Thanks to Gaia DR2, \citet{2019A&A...621A..38G} compiled an all-sky catalogue of $39\,800$ hot subdwarf star candidates by using the means of colour, absolute magnitude and reduced proper motion cuts. Except for the Galactic plane, the catalogue is nearly complete up to about $1.5\,{\rm kpc}$. The upper right panel of Figure$\,$\ref{fig:fig6} illustrates the absolute Gaia G magnitude $M_{\rm G}=G+5\log(D)-10-A_{G}$ versus distance D (in pc). In order to avoid contamination due to WDs at the faint limit, we restricted the sample to objects with $-0.65\le M_{\rm G}<-0.5$. The lower left panel of Figure$\,$\ref{fig:fig6} displays the distribution function of $M_{\rm G}$ for objects that lie in three distance intervals, respectively. The last two intervals show quite similar distribution functions and the Kolmogorov-Smirnov (K.S) test gives a $P$ value of 0.99. Therefore, the objects with $500<D<1\,500$ in Gaia DR2 are expected to be volume complete. The lower right panel of Figure$\,$\ref{fig:fig6} shows comparisons of the distribution function of $M_{\rm G}$ for hot subdwarf stars with $500<D<1\,500$ in Gaia DR2, objects in LAMOST DR5 \citep{2019ApJ...881....7L}, DR6 and DR7 \citep{2020ApJ...889..117L}. We consider the sample of hot subdwarfs in LAMOST DR5 to be complete. In the lower left panel of Figure$\,$\ref{fig:fig5} we give the volume selection effect corrected fractional distributions of the four hot subdwarf helium groups for the halo, thick disk and thin disk populations. The influence of {\bf the volume selection effect} on the results is estimated to be less than $5\%$ of the number of stars within a group. We also present the volume and RV variability selection effect corrected fractional distributions of the four hot subdwarf helium groups for the halo, thick disk and thin disk populations in the lower left panel of Figure$\,$\ref{fig:fig5}. A total impact of these two effects is less than $8\%$ of the number of stars within each group. We can see that the overall tendency of the fractional distributions of the four hot subdwarf helium groups in the halo, thin disk and thick disk from DR6 and DR7 are consistent with the findings reported by \cite{2019ApJ...881....7L} based on LAMOST DR5.

\section{Conclusions} \label{sec:conc}
To supplement our previous work \citep{2019ApJ...881....7L}, we explored the kinematics of 182 single-lined hot subdwarf stars selected by \citet{2020ApJ...889..117L} in Gaia DR2 with spectra from LAMOST DR6 and DR7. Making use of the parallaxes and proper motions of Gaia DR2 and the radial velocities measured from LAMOST spectra, we computed the Galactic space positions, Galactic velocity components and Galactic orbits. Following our previous work \citep{2019ApJ...881....7L}, these stars were classified into four groups based on their helium abundances. From the kinematic properties of the four hot subdwarf helium groups the following conclusions can be drawn:
\begin{enumerate}
  \item
  The space distributions show that the space density of He-rich stars have a larger dispersion than the groups of He-deficient stats from the thin disk to halo. The latter two groups do not show any obvious differences in space distribution, but the former two groups exhibit a noticeable difference around $|Z|=1.5$ kpc where the star density of He-rich stars with $\log(y)>0$ is far higher than that of stars with $-1\le\log(y)<0$.
  As described in \citet{2019ApJ...881....7L}, the space distribution differences indicate that He-rich and He-deficient stars likely originate from different formation channels.
  \item
  Likewise, the $U-V$ velocity diagram and the kinetic energy $2E_{\rm kin}/m=U^{2}+V^{2}+W^{2}$ versus rotational velocity ($V$) diagram demonstrate that He-deficient stars tend to group around the LSR, while He-rich stars are widely scattered in the whole parameter space. He-rich stars with $\log(y)>0$ have higher proportion than stars with $-1\le\log(y)<0$. In addition, He-rich stars with $\log(y)>0$ display the largest standard deviation of the Galactic velocity components and orbital parameters, while He-rich stars with $-1\le\log(y)<0$ represent the second largest standard deviation. The two groups of He-deficient stars with $\log(y)<-2.2$ display a similar value of standard deviation. These results also support that these four hot subdwarf helium groups are likely to origin from different formation channels \citep{2019ApJ...881....7L}.
  \item
  We have also presented a kinematic population classification of the four hot subdwarf helium groups based on their positions in the $U-V$ velocity diagram, $J_{Z}-e$ diagram and their Galactic orbits.
  The relative contributions of the four hot subdwarf helium groups to the halo, thin disk and thick disk can be largely matched with the results derived from LAMOST DR5 \citep{2019ApJ...881....7L}, which appears to support the predictions of binary population synthesis \citep{2003MNRAS.341..669H,2008A&A...484L..31H}. He-deficient stars with $-2.2\le\log(y)<-1$ likely origin from the stable RLOF channel, He-deficient stars with $\log(y)<-2.2$ from the CE ejection channel and He-rich stars with $\log(y)\ge0$ from the merger channel of double HeWDs. As shown by \cite{2019ApJ...881....7L}, the fraction of He-rich hot subdwarf stars with $-1\le\log(y)<0$ in the thin disk and halo is higher than in the thick disk, which suggests that these stars may have different formation channels in the thin disk and the halo.
\end{enumerate}

\acknowledgments

We would like to thank the anonymous reviewer for his/her comments, which significantly improved the paper.
The research presented here is supported by
the National Natural Science Foundation of China under
grant no. U1731111 and the Fundamental Research Fund of China West Normal University under grant no.17YC511.
P.N. acknowledges support from the Grant Agency of the Czech Republic (GA\v{C}R 18-20083S).
Guoshoujing Telescope (the Large Sky Area Multi-Object Fiber Spectroscopic
Telescope LAMOST) is a National Major Scientific
Project built by the Chinese Academy of Sciences.
Funding for the project has been provided by the National
Development and Reform Commission. LAMOST
is operated and managed by the National Astronomical
Observatories, Chinese Academy of Sciences.
This work has made use of data from the European Space Agency (ESA) mission
{\it Gaia} (\url{https://www.cosmos.esa.int/gaia}), processed by the {\it Gaia}
Data Processing and Analysis Consortium (DPAC,
\url{https://www.cosmos.esa.int/web/gaia/dpac/consortium}). Funding for the DPAC
has been provided by national institutions, in particular the institutions
participating in the {\it Gaia} Multilateral Agreement.
This research has used the services of \mbox{\url{www.Astroserver.org}}.
\software{\,astropy \citep{2013A&A...558A..33A, 2018AJ....156..123A}, \,TOPCAT (v4.6; \citealt{2005ASPC..347...29T,2018arXiv181109480T}), \,galpy \citep{2015ApJS..216...29B}}.

\begin{deluxetable*}{rlll}
\tablecaption{Atmospheric parameters, space positions, orbital parameters and galactic velocities for 182 single-lined hot subdwarf stars observed in Gaia DR2 and LAMOST DR6 and DR7. \label{tab:tab1}}
\tablewidth{0pt}
\tabletypesize{\scriptsize}
\tablehead{
\colhead{Num} & \colhead{Label}    & \colhead{Explanations}
}
\startdata
1  &   LAMOST          & LAMOST target\\
2  &      $RAdeg$      & Barycentric Right Ascension (J2000) \tablenotemark{(1)}\\
3  &      $DEdeg$      & Barycentric Declination (J2000) \tablenotemark{(1)}\\
4  &    $T_{\rm eff}$  & Stellar effective temperature \tablenotemark{(2)}\\
5  & $e\_T_{\rm eff}$  & Standard error in $T_{\rm eff}$ \\
6  &         $\log g$  & Stellar surface gravity \tablenotemark{(2)}\\
7  &      $e\_\log g$  & Standard error of Stellar surface gravity \\
8  &        $\log(y)$  & Stellar surface He abundance $y=n{\rm He}/n{\rm H}$ \tablenotemark{(2)}\\
9  &     $e\_\log(y)$  & Standard error in $\log(y)$ \\
10 &             type  & Spectra type \tablenotemark{(2)} \\
11 &           $pmRA$  & Proper motion in RA \\
12 &        $e\_pmRA$  & Standard error $pmRA$ \\
13 &           $pmDE$  & Proper motion in DE \\
14 &        $e\_pmDE$  & Standard error in $pmDE$\\
15 &              $D$  & Gaia DR2 stellar distance\\
16 &           $e\_D$  & Standard error in stellar distance\\
17 &           $RVel$  & Radial velocity from LAMOST spectra\\
18 &        $e\_RVel$  & Standard error in radial velocity\\
19 &              $X$  & Galactic position towards Galactic center\\
20 &           $e\_X$  & Standard error in $X$\\
21 &              $Y$  & Galactic position along Galactic rotation\\
22 &           $e\_Y$  & Standard error of $Y$\\
23 &              $Z$  & Galactic position towards north Galactic pole\\
24 &           $e\_Z$  & Standard error of $Z$\\
25 &              $U$  & Galactic radial velocity positive towards Galactic center\\
26 &           $e\_U$  & Standard error in $U$ \\
27 &              $V$  & Galactic rotational velocity along Galactic rotation \\
28 &           $e\_V$  & Standard error in $V$\\
29 &              $W$  & Galactic velocity towards north Galactic pole\\
30 &           $e\_W$  & Standard error in $W$ \\
31 &         $R_{ap}$  & Apocenter radius \tablenotemark{(3)} \\
32 &      $e\_R_{ap}$  & Standard error in $R_{ap}$\\
33 &       $R_{peri}$  & Pericenter radius \tablenotemark{(3)}\\
34 &    $e\_R_{peri}$  & Standard error in $R_{peri}$\\
35 &    $z_{\rm max}$  & Maximum vertical height \tablenotemark{(3)}\\
36 & $e\_z_{\rm max}$  & Standard error in $z_{\rm max}$\\
37 &              $e$  & Eccentricity \tablenotemark{(3)}\\
38 &           $e\_e$  & Standard error in $e$\\
39 &      $J_{\rm z}$  & Z$-$component of angular momentum \tablenotemark{(3)}\\
40 &   $e\_J_{\rm z}$  & Standard error in $J_{\rm z}$ \\
41 &      $z_{\rm n}$  & Normalised z-extent of the orbit \tablenotemark{(3)}\\
42 &   $e\_z_{\rm n}$  & Standard error in $z_{\rm n}$\\
43 &     Pops          & Population classification \tablenotemark{(4)}\\
44 & $P_{\rm TH}$      & probability in thin disk\\
45 & $P_{\rm TK}$      & probability in thick disk\\
46 & $P_{\rm H}$       & probability in halo\\
\enddata
\tablenotetext{(1)}{At Epock 2000.0 (ICRS).}
\tablenotetext{(2)}{From \cite{2020ApJ...889..117L}.}
\tablenotetext{(3)}{Form the numerical orbit integration.}
\tablenotetext{(4)}{H$=$Halo; TK$=$thick disk; TH$=$thin disk.}
\tablecomments{The full table can be found in the online version of the paper.}
\end{deluxetable*}

%\begin{longrotatetable}
\begin{longrotatetable}
\begin{deluxetable*}{lclclclclclclclclclclclc}
\tablecaption{Mean values and standard deviations of the Galactic velocities and the Galactic orbital parameters: eccentricity ($e$), normalised z-extent ($z_{\rm n}$),  maximum vertical amplitude ($z_{\rm max}$), apocentre ($R_{\rm ap}$) and pericentre  ($R_{\rm peri}$) for the four hot subdwarf helium groups. \label{tab:tab2}}
\tablewidth{1500pt}
%\tabletypesize{\scriptsize}
\tablehead{
\colhead{Subsample} & \colhead{N} &
\colhead{$\bar{U}$} & \colhead{$\sigma_{U}$} &
\colhead{$\bar{V}$} & \colhead{$\sigma_{V}$} &
\colhead{$\bar{W}$} & \colhead{$\sigma_{W}$} &
\colhead{$\overline{U^{2}+V^{2}+W^{2}}$} & \colhead{$\sigma_{U^{2}+V^{2}+W^{2}}$} &
\colhead{$\overline{e}$}           & \colhead{$\sigma_{e}$}           &
\colhead{$\overline{z_{\rm n}}$}   & \colhead{$\sigma_{z_{\rm n}}$}   &
\colhead{$\overline{z_{\rm max}}$} & \colhead{$\sigma_{z_{\rm max}}$} &
\colhead{$\overline{R_{ap}}$}      & \colhead{$\sigma_{R_{ap}}$}       &
\colhead{$\overline{R_{peri}}$}    & \colhead{$\sigma_{R_{peri}}$}
}
\startdata
All stars           &182& 30& 62& 203& 35&  0& 36& 46$\,$458& 16$\,$817& 0.23& 0.13& 0.12& 0.08& 1.14& 0.70&  9.94& 1.84& 5.98& 2.45\\
$\log(y)\ge0$       & 20& 39& 76& 148& 82&  4& 60& 42$\,$938& 20$\,$150& 0.41& 0.32& 0.29& 0.22& 2.42& 1.75& 10.17& 2.46& 4.79& 3.22\\
$-1\le\log(y)<0$    & 12&  3& 43& 204& 44& -2& 49& 47$\,$907& 19$\,$653& 0.27& 0.20& 0.22& 0.23& 2.38& 2.37& 10.96& 1.51& 6.86& 2.99\\
$-2.2\le\log(y)<-1$ & 57& 35& 57& 205& 33& -1& 31& 46$\,$568& 12$\,$457& 0.20& 0.10& 0.12& 0.08& 0.99& 0.53&  9.64& 1.46& 6.13& 2.15\\
$\log(y)<-2.2$      & 89& 29& 63& 203& 35&  3& 35& 46$\,$962& 16$\,$640& 0.24& 0.14& 0.12& 0.07& 1.17& 0.73&  9.94& 1.89& 6.02& 2.25\\
\enddata
\end{deluxetable*}
\end{longrotatetable}

\begin{deluxetable*}{lclclclclc}
\tablecaption{Population classification and relative contributions of the four hot subdwarf helium groups.\label{tab:tab3}}
\tablewidth{0pt}
\tablehead{
\colhead{Subsample}  &
\colhead{N}          &
\colhead{Thin Disk}  &
\colhead{Thick Disk} &
\colhead{Halo}
}
\startdata
All stars           & 182  &  83  & 73  & 26  \\
$\log(y)\ge0$       & 19   &  5   &  8   & 6 \\
$-1\le\log(y)<0$    & 13   &  7   &  2   & 4  \\
$-2.2\le\log(y)<-1$ & 60   & 32   &  20  & 8  \\
$\log(y)<-2.2$      & 90   & 39   &  43  & 8  \\
\enddata
\end{deluxetable*}

\end{document}